Comments on
"Observation of Long-Range, Near-Side Angular
Correlations in Proton-Proton Collisions at the LHC" by
the CMS collaboration [1].

Michael J. Tannenbaum[a] and Richard M. Weiner[b]

In a recent preprint [1] the CMS collaboration presented results showing an "unexpected" long - range, near-side two-particle correlation in pp reactions at the Cern LHC, "reminiscent of correlations seen in relativistic heavy ion data" and which cannot be explained by current event generators. The authors conclude that the physical origin of their observation is not yet understood.

It is the purpose of this note to point out that this observation is in line with previous observations in *particle physics* at large transverse momenta and/or high multiplicities but lower energies and which were interpreted [2], [3] as possible evidence for quark-gluon plasma (QGP), even before this state of matter has been observed in heavy ion reactions. The fact that a correlation strikingly similar to that reported in [1] was observed in a p-p experiment at the CERN ISR (cf. Ref. 4, Fig. 4) suggests that the same mechanism may be responsible for the results of Ref. 1 and explains the analogy with nucleus-nucleus reactions at RHIC alluded to in that reference.

The evidence for QGP mentioned above is based in part on the success of the hydrodynamical model of Landau in describing multiparticle production in strong p-p and p-nucleus reactions: hydrdynamics assumes *local equilibrium*, which in turn presupposes a large number of degrees of freedom. QGP is the only known mechanism which can provide this large number.

It is also based on the particular form of the *equation of state*, which enters the hydrodynamical approach and which is characteristic for QGP.

---

[a] Physics Department, Brookhaven National Laboratory, Upton, New York, USA
[b] Laboratoire de Physique Théorique, Univ. Paris XI, France and Physics Department, University of Marburg, Germany

A review of the relationship between QGP and particle reactions can be found in Refs.5 and 6. In these references the universality of the hadronization process as well as the evidence for QGP from CERN and RHIC heavy ion reactions are also discussed. In view of these remarks the results of Ref. 1 appear less surprising[c] and can be considered as further evidence for QGP in particle reactions. The fact that this refers to energies two orders of magnitude higher than those previously available strengthens considerably this line of argumentation.

However in order to truly understand whether these and previous effects are evidence for the QGP in p-p reactions, several other features of the QGP observed in A+A collisions at RHIC must be investigated. These include radial flow, i.e. the increase of $<p_T>$ for heavy particles such as protons with increasing associated multiplicity. More importantly, a key test for a QGP in A+A collisions [7] was the suppression of hadrons at large transverse momentum ($p_T$) due to the interaction of the outgoing partons from hard-scattering with the produced medium. A manifestation of QGP in p-p collisions would be the suppression of high $p_T$ particles at high multiplicity. Obviously there must be some effect due to energy conservation, but comparing production at mid-rapidity in p-p collisions of direct photons (which don't interact with the medium) to $\pi_0$ production, as done for A+A collisions in PHENIX [8], will show whether any suppression is due to conservation of energy or is a medium effect. Observation of both these features in high-multiplicity p-p collisions would be strong evidence for a QGP.

References

[1]The CMS collaboration, Observation of Long-Range, Near-Side Angular Correlations in Proton-Proton Collisions at the LHC, **arXiv:1009.4122**.
[2] E. M. Friedlander and R. M. Weiner, Evidence from Very Large Transverse Momenta of a Change with Temperature of Velocity of Sound in Hadronic Matter, Phys. Rev. Letters, 43, (1979) 15. In this reference extrapolations for LHC energies are made.
[3] G. N. Fowler, E. M. Friedlander, R. M. Weiner and G. Wilk, Possible Manifestation of Quark-Gluon Plasma in Multiplicity Distributions from High Energy Reactions, Phys. Rev. Letters 57, 2119 (1986).

---

[c] As emphasized by the very title of Ref. 5 apparent surprises in this domain were reported also previously.